\input phyzzx.tex
\tolerance=1000
\voffset=-0.0cm
\hoffset=0.7cm
\sequentialequations

\def\t1{{\tilde 1}}

\def\rn{Reissner--Nordstrom~}

\def\t{\theta}
\def\S{Schwarzschild~}
\def\p{^\prime}

\REF{\HOL}{G. 't Hooft, [arXiv:gr-qc/9310026]; L. Susskind, J. Math. Phys. {\bf 36} (1995) 6377, [arXiv:hep-th/9409089]; R. Bousso, Rev. Mod. Phys. {\bf 74} (2002) 825, [arXiv:hep-th/0203101].}
\REF{\LAST}{E. Halyo, [arXiv:1706.07428]; [arXiv:1805.06079].}
\REF{\ADS}{J. Maldacena, Adv. Theor. Math. Phys. {\bf 2} (1998) 231, [arXiv:hep-th/9711200]; S. Gubser, I. Klebanov and A. Polyakov, Phys. Lett. {\bf B428} (1998) 105, [arXiv:hep-th/9802109]; E. Witten, Adv. Theor. Math. Phys. {\bf 2} (1998) 253, [arXiv:hep-th/9802150].}
\REF{\GRU}{ D. Grumiller and R. McNees, JHEP {\bf 0704} (2007) 074, [arXiv:hep-th/0703230].}
\REF{\REV}{M. Rangamani and T. Takayanagi, Lect.Notes Phys. {\bf 931} (2017) 1, [arXiv:1609.01287].}
\REF{\SEN}{A. Sen, Entropy {\bf 13} (2011) 1305, [arXiv:1101.4254].}
\REF{\ENT}{S. Ryu and T. Takayanagi, Phys. Rev. Lett. {\bf 96} (2006) 181602, [arXiv:hep-th/0603001; JHEP {\bf08} (2006) 045,
[arXiv:hep-th/0605073]; T. Nishioka, S. Ryu and T. Takayanagi, JPhys. {\bf A42} (2009) 504008, [arXiv:0905.0932].}
\REF{\TAK}{T. Azeyanagi, T. Nishioka and T. Takayanagi, Phys.Rev. {\bf D77} (2008) 064005, [arXiv:0710.2956].}
\REF{\VAN}{B. Czech, J. L. Karczmarek, F. Nogueira and M. Van Raamsdonk, Class. Quant. Grav. {\bf 29} (2012) 235025, [arXiv:1206.1323].}
\REF{\JT}{R. Jackiw, Nucl. Phys. {\bf B252} (1985) 343; C. Teitelboim, Phys. Lett. {\bf B126} (1983) 41.}
\REF{\MIG}{M. Cadoni and S. Mignemi, Phys. Rev. {\bf D51} (1995) 4319, [arXiv:hep-th/9410041].}
\REF{\RIN}{M. Parikh and P. Samantray, [arXiv:1211.7370].}
\REF{\NHO}{H. K. Kunduri, J. Lucetti and H. S. Reall, Class. Quant. Grav. {\bf 24} (2007) 4169, [arXiv:0705.4214];
P. Figueras, H. K. Kunduri, J. Lucetti and M. Rangamani, Phys. Rev. {\bf D78} (2008) 044042, [arXiv:0803.2998].}
\REF{\DAB}{A. Dabholkar, A. Sen and S. P. Trivedi, JHEP {\bf 0701} (2007) 096, [arXiv:hep-th/0611143].}
\REF{\SYK}{A. Kitaev, http://online.kitp.ucsb.edu/online/entangled15/kitaev/, http:// \hfill
online.kitp.ucsb.edu/online/entangled15/kitaev2/,
J. Madacena and D. Stanford, Phys.Rev. {\bf D94} (2016) no.10, 106002, [arXiv:1604.07818]; J. Madacena, D. Stanford and Z. Yang,
PTEP 2016 (2016) no.12, 12C104, [arXiv:1606.01857].}
\REF{\CARL}{S. Carlip, Phys. Rev. Lett. {\bf 82} (1999) 2828, [arXiv:hep-th.9812013]; Class. Quant. Grav. {\bf 16} (1999) 3327,
[arXiv:gr-qc/9906126].}
\REF{\SOL}{S. Solodukhin, Phys. Lett. {\bf B454} (1999) 213, [arXiv:hep-th/9812056].}
\REF{\EDI}{E. Halyo, [arXiv:1502.01979]; [arXiv:1503.07808]; [arXiv:1506.05016]; [arXiv:1606.00792].}
\REF{\PGB}{E. Halyo, [arXiv:1809.10672].}

\singlespace
\pagenumber=0
\normalspace
\medskip
\bigskip
\titlestyle{\bf{Computing Black Hole Entropy at Infinity}}
\smallskip
\author{Freya Edholm$^1${\footnote*{email: fedholm@my.smccd.edu}} and Edi Halyo$^2${\footnote 
\dagger{email: halyo@stanford.edu}}}
\smallskip
\centerline {$^1$  947 Laurel Avenue} 
\centerline {San Mateo, CA 94401}
\smallskip
\smallskip
\centerline {$^2$ Department of Physics} 
\centerline{Stanford University} 
\centerline {Stanford, CA 94305}
\smallskip

\vskip 2 cm
\titlestyle{\bf Abstract}

We show that the entropy of asymptotically flat, nonextremal black holes can be computed at infinity. We provide a prescription for transforming these black holes to $AdS_2$ black holes with the same entropy by dimensional reduction to 2D and a Weyl transformation. We apply our prescription to Schwarzschild, 4D Reissner--Nordstrom and generic nonextremal black holes. In the transformed coordinates, the asymptotic 
regions contain a global $AdS_2$ whose entropy can be computed either as the holographic entanglement entropy or as the entanglement entropy of a pair of Rindler $AdS_2$ spaces in the thermofield double state. This precisely reproduces the entropy of the original nonextremal black holes.

\singlespace
\vskip 0.5cm
\endpage
\normalspace

\centerline{\bf 1. Introduction}
\medskip

The holographic principle states that the fundamental degrees of freedom that describe
gravity in a region are located on its boundary[\HOL]. As a result, it is universally believed that the degrees of freedom of a black hole are located on its horizon. It then follows that black hole entropy should be computed on the horizon. 
In this paper, following ref. [\LAST], we would like to argue that nonextremal (and asymptotically flat) black hole entropy can be computed at asymptotic infinity. At first, this does not seem possible since this region is flat. Clearly, 
black hole entropy cannot be computed at infinity using a coordinate system in which the metric is asymptotically flat. However, this becomes possible in a ``better" coordinate system, i.e. in the 2D dimensionally reduced black hole metric after a proper Weyl transformation. Under the dimensional reduction and Weyl transformation, black hole thermodynamics remains invariant; in particular, entropy remains the same. As a result, we can transform a 
D--dimensional nonextremal black hole to an $AdS_2$ black hole with the same entropy. Asymptotic infinity in the original coordinates corresponds to the global $AdS_2$ limit. Then, the entropy of global $AdS_2$ reproduces that of the original nonextremal black hole showing that black hole entropy can be computed asymptotically.

This result may seem less strange if we consider a spherical shell that collapses from infinity to form a black hole. It is natural to assume that initially the shell and its surroundings are in a pure but entangled state. Once the shell passes its \S radius and forms a black hole, its degrees of freedom are no longer available to us. Thus, we need to trace over these
and find that the black hole is described by a mixed state with nonvanishing entanglement entropy. It is 
well--known that, for a pure state, the entanglement entropy of a subsystem and that of its complement are the same.
Therefore, the entanglement entropy of the black hole is identical to that of the degrees of freedom at infinity with which it initially formed a pure state. This means that, in principle, black hole entropy can be computed at infinity as an entanglement entropy.

Computing black hole entropy at asymptotic infinity is also quite similar to what happens in the AdS/CFT correspondence[\ADS]. In AdS/CFT duality, an AdS black hole is described by a state on the boundary which is located at asymptotic infinity and not on the horizon of the black hole.
In our description, the asymptotic region can be considered a screen just like the AdS boundary on which the fundamental black hole degrees of freedom live. As mentioned above, this is only visible in the dimensionally reduced and Weyl transformed coordinates.

In this paper, we give a prescription for transforming any nonextremal, asymptotically flat black hole to an $AdS_2$ black hole. We then apply the prescription to D--dimensional \S black holes, the 4D \rn black hole and
generic nonextremal black holes. In each case, starting with a higher dimensional nonextremal black hole, we first dimensionally reduce the 
D--dimensional metric over the transverse $D-2$ directions, leaving only the t--r directional metric. In 2D, the transverse directions are parametrized by the dilaton field. This solution describes a dilatonic black hole with the same thermodynamics as the original
nonextremal one. We then Weyl transform this dilatonic black hole into an $AdS_2$ black hole and show that the asymptotic limit in the original black hole coordinates is global $AdS_2$ in each case. We compute the entropy of global $AdS_2$ by two different methods. First, we compute the holograpic entanglement of global $AdS_2$.
Alternatively, we calculate the entanglement entropy of a pair of Rindler $AdS_2$ space--times in a thermofield double state that is equivalent to global $AdS_2$. Rindler $AdS_2$ is identical to an $AdS_2$ black hole with an $AdS_2$ radius. The entropy of this black hole is the entanglement entropy of the Rindler $AdS_2$ in a thermofield double state. Both computations reproduce the correct entropy for the original nonextremal black hole. 

This paper is organized as follows. In the next section we review 2D dilatonic gravity, its black holes and the Weyl transformations
that preserve their thermodynamics. We then give our prescription for transforming D--dimensional nonextremal black holes
to $AdS_2$ black holes. In sections 3, 4, and 5 we apply our prescription to D--dimensional \S black holes, the 4D \rn black hole and
generic nonextremal black holes respectively. In section 6, we show that the asymptotics of these black holes, i.e. global $AdS_2$
has an entropy that reproduces the entropy of the original nonextremal black hole. Section 7 contains a discussion of our results and our conclusions.

\bigskip
\centerline{\bf 2. Transformation of Nonextremal Black Holes to $AdS_2$ Black Holes}
\medskip

We begin with a brief review of 2D dilatonic gravity, its black hole solutions and Weyl transformations in this theory.
The generic 2D dilatonic gravity action is given by[\GRU]
$$I={1 \over 2} \int d^2x \sqrt{-g} \left[\phi R- U(\phi) (\nabla\phi)^2+ V(\phi) \right] \quad, \eqno(1)$$
where $\phi$ is the dilaton and $U(\phi)$ and $V(\phi)$ are the kinetic and potential functions respectively. 
This theory has black hole solutions given by the metric and a dilaton profile
$$ds^2=-f(r)dt^2+f(r)^{-1} dr^2 \qquad \phi=\phi(r) \quad, \eqno(2)$$
where
$$f(r)=e^{Q(\phi)} (\omega(\phi)-2M) \qquad {{\partial \phi} \over {\partial r}}= e^{-Q(\phi)} \quad. \eqno(3)$$
Here $M$ is the black hole mass and $Q(\phi)$ and $\omega(\phi)$ are defined by
$$Q(\phi)=\int^{\phi} d{\bar \phi}~ U({\bar \phi}) \qquad \omega(\phi)=\int^{\phi} d{\bar \phi}~V({\bar \phi})e^{Q({\bar \phi})}
\quad. \eqno(4)$$
The black hole horizon is at $\phi_h$ which satisfies
$$\omega(\phi_h)=2M \quad. \eqno(5)$$
The temperature and entropy of the black hole are given by
$$T_H={1 \over {4 \pi}}{d\omega \over d\phi}(\phi_h) \qquad \qquad S_{BH}=2 \pi \phi_h \quad. \eqno(6)$$
With the normalization of the action in eq. (1), the two dimensional Newton constant is determined by $\phi_h$ to be
$$G_2={1 \over {8 \pi \phi_h}} \quad. \eqno(7)$$
Under Weyl transformations of the form
$$g_{\mu \nu} \to {g\p}_{\mu \nu}=e^{-2 \sigma(\phi)}g_{\mu \nu} \quad, \eqno(8)$$
the action in eq. (1) becomes
$$I={1 \over 2} \int dt dr^{\prime} \sqrt{-{g\p}} \left[\phi {R\p}- {U\p}(\phi) (\nabla\phi)^2+ {V\p}(\phi) \right] \quad, \eqno(9)$$
where the transformed kinetic and potential functions are given by
$${U\p}(\phi)=U(\phi)-2 {{d\sigma(\phi)} \over {d\phi}} \qquad {V\p}(\phi)=e^{2 \sigma(\phi)} V(\phi) \quad. \eqno(10)$$
As a result, we find that
$${Q\p}(\phi)=Q(\phi)-2 \sigma(\phi) \qquad {\omega\p}(\phi)=\omega(\phi) \quad. \eqno(11)$$
Under a Weyl transformation, the form of the black hole solution given by eqs. (2) and (3) remains invariant with $Q$ replaced by ${Q\p}$ and $r$ replaced by $r^{\prime}$ which is determined by
$\partial_{r^{\prime}}=e^{2 Q^{\prime}(\phi)} \partial_r$. 
It is crucial that $\omega(\phi)$ is invariant under Weyl transformations so the thermodynamics of 2D dilatonic black holes is Weyl invariant. Thus, we can use a Weyl transformation to transform the 2D dilatonic black hole described above into $AdS_2$ black holes with the same thermodynamics.

We now return to our original problem, i.e. computing the entropy of D--dimensional nonextremal black holes with metrics that are generically of the form
$$ds^2=-f(r)dt^2+f(r)^{-1} dr^2 +r^2 d\Omega_{D-2}^2 \quad. \eqno(12)$$
Here, $f(r)$ is a function with $f(r_0)=0$ and $f^{\prime}(r_0) \not =0$ where $r_0$ is the black hole radius.
In order to compute the entropy, we first dimensionally reduce the D--dimensional metric in eq. (12) over the transverse $S^{D-2}$
to 2D and obtain a dilatonic black hole.
We then Weyl transform this into an $AdS_2$ black hole with the same entropy. For a 2D theory obtained by dimensional reduction over $S^{D-2}$, $G_2=G_D/A_h$ and therefore
$$S= 2 \pi \phi_h={1 \over {4 G_2}}={A_h \over {4 G_D}} \quad. \eqno(13)$$ 
We see that both 2D black holes have the same entropy as the original nonextremal black hole.

We now describe
our prescription for dimensionally reducing and Weyl transforming any D--dimensional nonextremal black hole into an $AdS_2$ black hole with the same entropy. The steps are:

1. Consider only the t--r directions of the D--dimensional metric and dismiss the $D-2$ transverse directions. This amounts to the dimensional reduction of the original D--dimensional metric over the transverse directions ($S^{D-2}$) to 2D
at the price of introducing a dilaton field which parametrizes the size of the transverse sphere. The 2D action is given by eq. (1) where $U(\phi)$ and $V(\phi)$ are obtained from the original metric that is dimensionally reduced.

2. Determine $e^{Q(r)}$ and $\omega(r)$ by using 
$$f(r)=e^Q  (\omega-2M) \quad. \eqno(14)$$ 

3. Find the dilaton profile $\phi(r)$ by solving
$${{d \phi(r)} \over {dr}}=e^{-Q(r)} \quad. \eqno(15)$$
$\phi(r)$ together with the 2D metric describes a 2D dilatonic black hole with the same thermodynamics as the original nonextremal black hole.

4. Invert $\phi(r)$ to get $r(\phi)$ and obtain $e^{Q(\phi)}$ and $\omega(\phi)$.

5. Assume a Weyl transformation of the form $e^{-2 \sigma}=A \phi^{\alpha}$ under which $Q$ transforms according to eq. (11). Here $A$ and $\alpha$ are constants to be determined by the requirement that the Weyl transformation leads to an $AdS_2$ black hole. The Weyl transformed 2D action is given by (9) with the transformed ${U\p}(\phi)$ and ${V\p}(\phi)$ given by eq. (10).

6. Solve for $\phi(r^{\prime})$ using
$${{d \phi(r^{\prime})} \over {dr^{\prime}}}=e^{-Q^{\prime(\phi)}} \quad. \eqno(16)$$
This can be inverted to give $r^{\prime}(\phi)$. Here $r^{\prime}$ is the new radial variable that is related to $r$ by 
$dr/dr^{\prime}=e^{2 \sigma}=A \phi^{\alpha}$.

7. Demand that the first term in the transformed $f(r^{\prime})$ which is $A \phi^{\alpha}$ reproduces that of an $AdS_2$ black hole i.e. $r^{\prime 2}/K^2 \ell_P^2$ where
$K$ is a large and arbitrary numerical factor and $\ell_P$ is the Planck length. This fixes both $A(K)$ and $\alpha$ and therefore the Weyl transformation that takes the dilatonic black hole into the $AdS_2$ black hole. $K$ is a large numerical factor that represents the global scale symmetry left over after the Weyl transformation is fixed. It will be chosen to be large enough so that the $AdS_2$ radius, $K \ell_P$, is much larger than $\ell_P$ and therefore gravity is semiclassical.

8. The second term (or the other terms in more general metrics) in $f(r^{\prime})$ is proportional to $r^{\prime p}$ where $p \geq 3$. In the asymptotic $r \to \infty$ limit of the original coordinates which corresponds to $r^{\prime} \to 0$, the first term dominates and  we are left with the pure $AdS_2$ metric.


\bigskip
\centerline{\bf 3. Schwarzschild Black Holes and the Asymptotic $AdS_2$} 
\medskip

Let us now apply the above prescription to D--dimensional Schwarzschild black holes with metric
$$ds^2 =-\left(1-{r_0^{D-3} \over r^{D-3}}\right)dt^2+\left(1-{r_0^{D-3} \over r^{D-3}}\right)^{-1}dr^2+r^2d\Omega_{D-2} \quad, \eqno(17)$$   
where the black hole horizon is at
$$r_0^{D-3}={{16 \pi G_D M} \over {(D-2)A_{D-2}}} \qquad A_{D-2}={{2 \pi^{(D-1)/2}} \over {\Gamma((D-1)/2)}} \quad. \eqno(18)$$ 
Dimensionally reducing D--dimensional General Relativity to 2D results in 2D dilatonic gravity with the action in eq. (1) with
$$U(\phi)=-{{(D-3)} \over {(D-2)}} {1 \over \phi} \qquad V(\phi)=\left({{(D-2)A_{D-2}} \over {8 \pi G}}\right)^{2/(D-2)} {{(D-3)} \over {(D-2)}}\phi^{(D-4)/(D-2)} \quad. \eqno(19)$$
Using eq. (14) we find that
$$e^{Q(r)}={1 \over {\omega(r)}}={{8 \pi G_D} \over {(D-2)A_{D-2}r^{D-3}}} \quad. \eqno(20)$$
$\phi(r)$ is determined by eq. (15) to be
$$\phi(r)={{A_{D-2} r^{D-2}} \over {8 \pi G_D}} \quad. \eqno(21)$$
As expected the black hole entropy is given by the horizon value of the dilaton $S=2 \pi \phi(r_0)$. Using eq. (21), $e^Q(\phi)$ and $\omega(\phi)$ can be written as 
$$e^{Q(\phi)}={1 \over {\omega(\phi)}}={{\ell_P} \over 2} \left({{8 \pi} \over A_{D-2}} \right)^{1/(D-2)} \phi^{(3-D)/(D-2)} \quad, \eqno(22)$$
where $\ell_P=G_D^{1/(D-2)}$ is the Planck length. Assuming a Weyl transformation of the form 
$e^{-2 \sigma}=A \phi^{\alpha}$ and using eq. (16) we find $r^{\prime}$ as a function of $\phi$
$$r^{\prime}=A \left({{8 \pi G_D} \over A_{D-2}}\right)^{1/(D-2)}{1 \over {(D-2)(1/(D-2)+\alpha)}} \phi^{1/(D-2)+\alpha} \quad. \eqno(23)$$
Now, demanding that the first term in $f(r^{\prime})$ matches that in the $AdS_2$ black hole metric, i.e. 
$${r^{\prime 2} \over {K^2 \ell_P^2}}=e^{-2 \sigma}=A \phi(r^{\prime})^{\alpha} \quad, \eqno(24)$$
we obtain $\alpha=2/(2-D)$ and $A=K^2 (8 \pi/A_{D-2})^{-2/(D-2)}$.
As explained above we assume $K>>1$. The action after this Weyl transformation is given by eq. (9) with
$${U\p}(\phi)={{(5-D)} \over {(D-2)}} {1 \over \phi} \qquad {V\p}(\phi)= \phi  \left({{A_{D-2}} \over {8 \pi}}\right)^{4/(D-2)} \left({{(D-2)} \over G_D}\right)^{2/(D-2)} K^2 \quad. \eqno(25)$$
We note that ${V\p}(\phi) \prop \phi$ which is a sign that the theory becomes $AdS_2$ gravity which of course is consistent with our demand above to get an $AdS_2$ black hole.

After this Weyl transformation $f(r^{\prime})$ that defines the black hole metric becomes
$$f(r^{\prime})=\left({{r^{\prime 2}} \over {K^2 \ell_P^2}}- {{2M(D-2) A_{D-2}} \over {K^{2D-4} \ell_P^{D-2}}} r^{\prime (D-1)}
\right) \quad, \eqno(26)$$
where $r^{\prime}$ is determined by $dr/dr^{\prime}=e^{2 \sigma}=A \phi^{\alpha}$ which gives
$$r^{\prime}={{K^2 \ell_p^2} \over r} \quad. \eqno(27)$$
We find that $r$ and $r^{\prime}$ are basically reciprocals of each other. As a result, the asymptotic limit in the original \S 
coordinates, $r \to \infty$ corresponds to $r^{\prime} \to 0$. In this limit the metric described by eq. (26) becomes pure $AdS_2$.

More precisely, for generic $r^{\prime}$, the metric given by eq. (26) corresponds to an $S^{D-2}$ parametrized by $\phi(r^{\prime})$, fibered over an $AdS_2$ black hole. The $AdS_2$ and $S^{D-2}$ radii are $K \ell_P$ and $\sim K^2 \ell_P^2/r^{\prime}$ respectively.
For $K^{D-2}>>M \ell_P$, as we decrease $r^{\prime}$ we find that there is a range of $r^{\prime}$, i.e. 
$K \ell_P<< r^{\prime}<<K^2(\ell_P^{D-4}/M)^{1/(D-3)}$,
for which the metric is essentially $AdS_2$ since the $S^{D-2}$ radius is much smaller than that of $AdS_2$. For smaller 
$r^{\prime} \sim K \ell_P$, the $S^{D-2}$ radius becomes about the $AdS_2$ radius and the transverse directions cannot be neglected. For $r^{\prime}<<K \ell_P$ we get a large $S^{D-2}$ fibered over $AdS_2$. On the other hand, if $K^{D-2}$ is not much larger than $M \ell_P$ the pure $AdS_2$ range does not exist and when we take $r^{\prime} \to 0$ we pass from the $AdS_2$ black hole directly to 
$S^{D-2}$ fibration over $AdS_2$. In summary, in the limit $r^{\prime} \to 0$, we get a large and growing $S^{D-2}$ with radius $K^2 \ell_P^2/r^{\prime}$ fibered over a pure $AdS_2$ space--time with a fixed radius $K \ell_P$.


\bigskip
\centerline{\bf 4. The 4D Reissner--Nordstrom Black Hole and the Asymptotic $AdS_2$}
\medskip

Our prescription can be applied to any nonextremal black hole that is asymptotically flat. (Thus, it does not apply to black holes in $AdS$ or $dS$ space--times.) As an example, in this section, we consider
the 4D Reisnner--Nordstrom black hole which is a solution to 4D Einstein-Maxwell gravity. It is well-known that the near horizon geometry of this black hole is $AdS_2 \times S^2$.
In this section, we show that this black hole geometry is asymptotically $AdS_2$ (with an $S^2$ fibration) after a Weyl transformation. The \rn black hole metric is given by
$$ds^2=-f(r)dt^2+f(r)^{-1} dr^2+r^2 d\Omega^2 \qquad, \eqno(28)$$
where
$$f(r)=1-{{2GM} \over r}-{{q^2G} \over {2r^2}} \quad. \eqno(29)$$
We now follow the prescription described in section 2. Dimensionally reducing 4D Einstein-Maxwell gravity
to 2D we get an action of the form given by eq. (1) with
$$U(\phi)={1 \over {2 \phi}} \qquad V(\phi)={1 \over G}-{q^2 \over {4G \phi}} \quad \eqno(30)$$
From eqs. (30) and (14) we find
$$e^Q(r)={G \over r} \qquad \omega(r)={r \over G}-{{q^2} \over {2r}} \quad, \eqno(31)$$
which gives $\phi(r)=r^2/2G$. As expected, the black hole entropy is $S=2 \pi \phi_0$. In this case, $e^{Q(\phi)}$ and $\omega(\phi)$ are given by
$$e^{Q(\phi)}=\sqrt{{G \over {2 \phi}}} \qquad \omega(\phi)= \sqrt{{{2 \phi} \over G}}- q^2 \sqrt{{G \over {8 \phi}}} \quad. \eqno(32)$$  
Again we consider a Weyl transformation of the form $e^{-2 \sigma}=A \phi^{\alpha}$ and using eq. (9) we find the transformed action with
$${U\p}(\phi)={1 \over \phi} \qquad {V\p}(\phi)={2 \over {K^2 G}} \phi-{q^2 \over {2 K^2 G}} \quad. \eqno(33)$$
We see that again ${V\p(\phi)}$ is proportional to $\phi$ which is consistent with $AdS_2$ black hole solutions. In this case, the new radial coordinate is
$$r^{\prime}=\sqrt{{G \over 2}} {A \over {\alpha+1/2}} \phi^{\alpha+1/2} \quad. \eqno(34)$$
Demanding that eq. (24) holds so that the Weyl transformation results in an $AdS_2$ black hole we obtain $\alpha=-1$ and $A=K^2/2$.
After the Weyl transformation $f(r^{\prime})$ becomes
$$f(r^{\prime})={{r^{\prime 2}} \over {K^2 \ell_P^2}}-{{2M r^{\prime 3}} \over {K^4 \ell_P^2}} -{{q^2 r^{\prime 4}} \over {2K^6 \ell_P^4}} \quad. \eqno(35)$$
We see that as expected the first term dominates in the asymptotic limit $r^{\prime} \to 0$ and the metric behaves very similarly 
to the \S case discussed at the end of the previous section giving rise to a large $S^{D-2}$ fibered over pure $AdS_2$. 

It is intriguing that the asymptotics of the Reissner--Nordstrom black hole is Weyl equivalent to $AdS_2$ which also appears in the near horizon geometry. Since $AdS_2$ is described by a CFT, it would be interesting to find out whether this change in the geometry from asymptotic infinity to the near horizon region can be related to a renormalization flow in the CFT.

\bigskip
\centerline{\bf 5. Nonextremal Black Holes and the Asymptotic $AdS_2$}
\medskip

In this section, we show that our prescription applies to all $D \geq 4$ nonextremal black holes which are asymptotically flat. Consider a generic metric of the form given by eq. (12) with
$$f(r)=1-{{aM} \over {r^{D-3}}}-\sum_i b_ir^{-c_i} \quad, \eqno(36)$$
where $a$ and $b_i$ are constants of length dimension $D-2$ and $c_i$ respectively and $c_i \geq 1$. This metric describes any type of nonextremal black hole with any charge. In addition, it describes black p--branes in D dimensions with the $D-p$ replacing $D$.

Following our prescription we find
$$e^{Q(r)}={a \over {r^{D-3}}} \qquad \omega(r)={{r^{D-3}} \over a}-\sum_i {b_i \over a} r^{D-3-c_i} \quad. \eqno(37)$$
Using eq. (15) we solve for $\phi(r)$
$$\phi(r)={{r^{D-2}} \over {a(D-2)}} \quad. \eqno(38)$$
$e^{Q(\phi)}$ and $\omega(\phi)$ are given by
$$e^{Q(\phi)}=a[a(D-2)\phi]^{(3-D)/(D-2)}  \quad \eqno(39)$$
and 
$$\omega(\phi)={1 \over a}[a(D-2)\phi]^{(D-3)/(D-2)}-\sum_i{b_i \over a}
[a(D-2)\phi]^{(D-3-c_i)/(D-2)} \quad. \eqno(40)$$
As before, a Weyl transformation of the form $e^{-2 \sigma}=A \phi^{\alpha}$ leads to
$$r^{\prime}={{aA[a(D-2)]^{(3-D)/(D-2)}} \over {(\alpha+1-(D-3)/(D-2))}} \phi^{(\alpha+1-(D-3)/(D-2))} \quad.  \eqno(41)$$
The condition given by eq. (24) fixes $\alpha$ and $A$ 
$$\alpha={-2 \over {(D-2)}} \qquad A={{K^2 \ell_P^2 (D-2)^{(D-3)/(D-2)}} \over {a^{2/(D-2)}}} \quad. \eqno(42)$$
The transformed metric for the $AdS_2$ black hole is of the form in eq. (2) with
$$f(r^{\prime})={{r^{\prime 2}} \over {K^2 \ell_P^2}}- C r^{\prime (D-1)}-\sum_i B_i r^{\prime (2+c_i)} \quad, \eqno(43)$$
where $C$ and $B_i$ are dimensionful constants. We see that in the asymptotic limit $r^{\prime} \to 0$ the first term dominates
and again we get a large $S^{D-2}$ fibered over pure $AdS_2$.

\bigskip
\centerline{\bf 6. Nonextremal Black Hole Entropy from the Asymptotic $AdS_2$}
\medskip

We have shown that the $AdS_2$ black holes obtained in the previous section have the same thermodynamics as those of D--dimensional nonextremal black holes. Our main assumption is that if the thermodynamics of these two space--times is the same, then the microscopic entropy counting should also be same. We also saw that the asymptotic limit of these black holes is pure $AdS_2$ 
space--time in which all signs of the black holes disappeared. However, the information about the $AdS_2$ black holes, i.e. the location of the horizon, is still present in the 2D Newton constant given by eq. (7). In addition, any entanglement that may exist is still present since it is a UV effect. 
We now show that the entropy of the $AdS_2$ black holes (and therefore that of the original D--dimensional nonextremal black holes) can be obtained either as the holographic
entanglement entropy of global $AdS_2$ or as the entanglement entropy of a pair of Rindler $AdS_2$ spaces in the thermofield double state. For more details see refs. [\ADS,\REV].

In the asymptotic limit $r^{\prime} \to 0$ of the $AdS_2$ black hole metric in eq. (43) the black hole disappears. This is the IR limit in the bulk and by the IR/UV duality corresponds to the UV limit of the boundary theory. 
In this limit, the finite temperature effects that describe the black hole on the boundary are negligible and the black hole disappears just as it does in the bulk. 
For $r^{\prime} \to 0$ the $AdS_2$ black hole metric becomes
$$ds^2=-{r^{\prime 2} \over {K^2 \ell_P^2}} dt^2+ {{K^2 \ell_P^2} \over {r^{\prime 2}}} dr^{\prime 2} \quad. \eqno(44)$$
This is the metric of the Poincare patch of $AdS_2$ which through the coordinate transformation given by[\SEN]
$$r^{\prime} \pm t=tan{1 \over 2}\left[{1 \over 2}(\sigma \pm \tau) \pm {\pi \over 2} \right] \quad, \eqno(45)$$
becomes global $AdS_2$ described by the metric
$$ds^2=K^2 \ell_P^2 {{-d \tau^2+d \sigma^2} \over sin^2\sigma} \quad. \eqno(46)$$
Global $AdS_2$ has two disconnected (one dimensional) boundaries at $\sigma=0,\pi$.

The global $AdS_2$ vacuum is described by a pure but entangled state of the two theories that live on the two boundaries[\TAK].
If we are restricted to only one boundary, then we need to trace over the states of the second one. As a result, the global $AdS_2$ vacuum becomes a mixed state described by a density matrix $\rho$. The entanglement entropy of this state is
$$S_{ent}=-Tr(\rho log \rho) \quad. \eqno(47)$$
$S_{ent}$ can be computed by using the holographic entanglement entropy formula[\REV.\ENT] 
$$S_{ent}(A)= {Area(\Sigma_A) \over {4G_2}} \quad, \eqno(48)$$
where $Area(\Sigma_A)$ is the area of the codimension two minimal surface in the bulk, $\Sigma_A$, such that the boundaries of $A$ and $\Sigma_A$ coincide. In our case, $A$ is one of the pointlike boundaries of $AdS_2$ and thus the minimal surface is a point in the bulk with 
$Area(\Sigma_A)=1$.  Therefore, from eq. (48) we get[\TAK]
$$S_{ent}(AdS_2)={1 \over {4 G_2}}={A_h \over {4 G_D}} \quad. \eqno(49)$$
This is the correct entropy for a generic nonextremal black hole.
The holographic entanglement entropy of global $AdS_2$ can also be computed in more detail by using the methods of refs. [\LAST] which we do not reproduce here. 

An alternative way to compute the entropy of global $AdS_2$ is to note that it is also described by two copies of
Rindler $AdS_2$ spaces in the thermofield double state. In general, global $AdS_n$ can be described as an entangled state of two zero mass hyperbolic black holes where the
black hole radii are equal to the $AdS_n$ radius[\VAN]. When their masses vanish, these hyperbolic black holes simply describe Rindler $AdS_n$ spaces; i.e. $AdS_n$ spaces seen from a frame with acceleration $1/r_{AdS}$.  In our case, since the boundary of $AdS_2$ is one dimensional, the hyperbolic nature of the boundary is irrelevant. As a result, global $AdS_2$ is described by two entangled $AdS_2$ Rindler spaces in the thermofield double state. 
By holography, $AdS_2$ Rindler spaces are described by their boundary theories. Therefore, the $AdS_2$ vacuum corresponds to the thermofield double state of the two boundary theories.
Again, if we are restricted to only one boundary, then we need to trace over the states of the second one. As a result, the 
entangled state that describes the global $AdS_2$ vacuum becomes a mixed state. 

In order to compute the entanglement entropy of this state, we use the fact that $AdS_2$ Rindler spaces are just $AdS_2$ black holes with $r_0=r_{AdS}$. The entanglement entropy  
of the mixed state is then given by the entropy of the corresponding black hole.

Consider dilatonic $AdS_2$ gravity (i.e. the Jackiw--Teitelboim theory [\JT]) with the action
$$I={1 \over {2}} \int d^2x \phi \left(R+{2 \over L^2}\right) \quad, \eqno(50)$$
where the cosmological constant is given by $\Lambda=-2/L^2$. 
This theory has dilatonic black holes with the metric[\MIG]
$$ds^2= -\left({r^2 \over L^2}- 2 M L \right)dt^2+\left({r^2 \over L^2}- 2 M L \right)^{-1}dr^2
\quad, \eqno(51)$$
and the linear dilaton profile $\phi= r/8 \pi G_2 L$ where again the normalization of the action in eq. (50) has been taken into account. The black hole horizon is at $r_0=(2 M L^3)^{1/2}$. 

Now, consider an $AdS_2$ black hole with $M=1/2 L$. The metric then becomes 
$$ds^2= -\left({r^2 \over L^2}- 1 \right)dt^2+\left({r^2 \over L^2}-1 \right)^{-1}dr^2  \quad, \eqno(52)$$
which is a black hole with $r_0=L$. 
This is actually an $AdS_2$ Rindler space with an acceleration $a=1/L$; i.e. the horizon at $r_0=L$ is an acceleration horizon. 
In order to see this, consider the coordinate transformation
$\rho=\sqrt{r^2-L^2}$ that takes the metric in eq. (52) to[\RIN]
$$ds^2=-{\rho^2 \over L^2}dt^2+\left(1+{\rho^2 \over L^2} \right)^{-1} d\rho^2 \quad. \eqno(53)$$
For $\rho<<L$ the metric describes Rindler space with $a=1/L$ whereas for $\rho>>L$ it becomes that of the Poincare patch of $AdS_2$. 
The entanglement entropy of the two $AdS_2$ Rindler spaces is the entropy of the black hole in eq. (52) which is given by 
$$S_{BH}={r_0 \over {4G_2 L}}={1 \over {4 G_2}}={A_h \over {4 G_D}} \quad, \eqno(54)$$
which matches the entropy of the original D--dimensional nonextremal black hole.

\bigskip
\centerline{\bf 7. Conclusions and Discussion}
\medskip

In this paper, we related D--dimensional nonextremal black holes asymptotically to $AdS_2$ space--times. 
Dimensionally reducing D--dimensional nonextremal black holes on an $S^{D-2}$ gives rise to 2D dilatonic black holes. 
These can be Weyl transformed into $AdS_2$ black holes which, due to the
invariance of 2D black hole thermodynamics under Weyl transformations, have the same entropy. In the asymptotic limit, i.e. $r \to \infty$ in the original coordinates, these metrics reduce to global $AdS_2$ which has the same entropy as the original D--dimensional nonextremal black holes. Our results indicate that black hole entropy can be computed at asymptotic infinity. Of course, this is not visible in the original asymptotically flat metric but becomes apparent only in the dimensionally reduced and Weyl transformed coordinates.

At asymptotic infinity where $r^{\prime} \to 0$, the dilaton given by eq. (41) diverges. This is simply the sign that the transverse $S^{D-2}$ parametrized by the dilaton decompactifies. Thus, the asymptotic geometry is given by a large $S^{D-2}$ fibered over
$AdS_2$ instead of flat space--time. Black hole entropy is carried by the $AdS_2$ factor and is not given by the area of the asymptotic $S^{D-2}$ which is much larger. This geometry can also be seen as global $AdS_2$ with two large $S^{D-2}$s on its boundaries. The theory that lives on the two $S^{D-2}$s, which are effectively parametrized by the $AdS_2$ boundary theory, are entangled. The entanglement entropy is given by that of global $AdS_2$ or two entangled Rindler $AdS_2$ space--times.

It is well--known that the near horizon geometries of extremal black holes contain an $AdS_2$ factor[\NHO] which is the origin of their entropies[\DAB]. In this paper, we found that generic nonextremal black holes, with a near horizon geometry that is Rindler space, have an $AdS_2$ factor which appears in their Weyl transformed asymptotic geometries and reproduces the correct black hole entropy. Thus, it seems that $AdS_2$ is the universal origin of both extremal and nonextremal black hole entropy. 

Even though we found that the entropy of nonextremal black holes is given by the 
holographic entanglement entropy of global $AdS_2$ or the entanglement entropy of two $AdS_2$ Rindler spaces in the thermofield double state, we do not have a clear idea about the nature of the black hole degrees of freedom that are counted. 
Following the AdS/CFT correspondence, we expect the boundary of $AdS_2$ to be described by a one dimensional CFT with only a time coordinate. The most promising description of the (near) $AdS_2$ boundary theory seems to be the SYK model[\SYK].  Unfortunately, the dual bulk description of the SYK model is problematic. Thus, the nature of the microscopic degrees of freedom counted by $AdS_2$ entropy remains an important open question.


It has been shown that the near horizon region of nonextremal black holes can be described by a horizon CFT that reproduces the black hole entropy[\CARL,\SOL,\EDI]. Nonextremal black hole entropy is also given by the pseudo Goldstone bosons (PGBs) of conformal symmetry that
lives in the very near horizon region. This symmetry is spontaneously broken by the Rindler vacuum and is anomalous. As a result, the low energy black hole physics is described by the PGBs of conformal symmetry[\PGB]. Their Schwarzian action is completely fixed by the symmetry breaking mechanism, i.e. conformal symmetry broken down to $SL(2)$. It is intriguing that the same Schwarzian action appears in the description
(near) $AdS_2$ space--times. Thus, we find that the near horizon physics of nonextremal black holes and their (Weyl transformed) asymptotic description are the same which leads more credence to the results in this paper.


\bigskip
\centerline{\bf Acknowledgments}

E.Halyo would like to thank the Stanford Institute for Theoretical Physics for hospitality.

\vfill

\refout

\end
\bye